\documentclass[paper]{geophysics}

\usepackage{booktabs}
\usepackage{multirow}
\usepackage{array}
\usepackage{xcolor}
\usepackage{amsmath, bm}
\usepackage[normalem]{ulem}
\usepackage{soul}

\renewcommand{\sout}[1]{\unskip}

\usepackage{draftwatermark}
\SetWatermarkText{Pre-print}
\SetWatermarkScale{2.1}

\begin{document}

\title{Experiences with Distributed Acoustic Sensing using both straight and helically wound fibers in surface-deployed cables - a case history in Groningen, The Netherlands}

\renewcommand{\thefootnote}{\fnsymbol{footnote}} 


\address{
\footnotemark[1]Department of Geoscience and Engineering, Delft University of Technology, \\
Stevinweg 1, \\
2628 CN Delft, The Netherlands \\
\footnotemark[2]Department of Geoscience and Engineering, Delft University of Technology, \\
Stevinweg 1, \\
2628 CN Delft, The Netherlands}

\author{Musab Al Hasani\footnotemark[1] and Guy Drijkoningen\footnotemark[2]}

\footer{Pre-print}
\lefthead{Al Hasani \& Drijkoningen}


\begin{abstract}
Distributed Acoustic Sensing (DAS) has been limited in its use for surface-seismic reflection measurements, due to the fiber’s decreased broadside sensitivity when the fiber is deployed horizontally. {Deploying the fiber in a helically wound fashion} \sout{Winding the fiber in a helically wound fashion} has the promise of being more sensitive to broadside waves (e.g. P-wave reflections) and less sensitive to surface waves than straight fiber. We examine these claims by burying a set of straight fibers (SF) and helically wound fibers (HWF) with different wrapping angles, using standard and engineered fibers. These fibers were buried in a 2 m deep trench in a farmland in the province of Groningen \sout{province} in the Netherlands. They are linked up to two interrogating systems and an electrically driven vibrator was used as a seismic source. We also deployed a 3-C geophone system for reference purposes.  We observe in our field data that using HWF has a destructive effect on the surface-wave amplitudes. Our data confirmed \sout{We also confirmed by our data} the effect of the wrapping angle on the polarity of the surface-wave arrival and the dampening effect of the helical winding, both behaving in quite a predictable fashion. Apart from the effect of the wrapping angle, the different design choices, e.g. cable filling and material type, did not show a significant effect on the amplitude of the signals. As for P-wave reflections, we observe that both engineered SF and HWF provide reflection images comparable \sout{comparable reflection images} to those obtained from the geophone data despite the straight fiber's decreased broadside sensitivity. A polarity reversal and an amplitude difference between the SF and HWF fibers are observed.
\sout{for which we provide a possible explanation via a COMSOL\textsuperscript{\textregistered} model.}
Finally, we show that the combined use of SF and HWF proved to be useful since SF showed better sensitivity in the shallower part and HWF in \sout{for} the deeper part.  
\end{abstract}
\section{Introduction}
Distributed Acoustic Sensing (DAS), sometimes called Distributed Vibration Sensing (DVS), has been widely adopted for a diverse number of applications. 
For borehole monitoring, DAS provides denser spatially sampled data than \sout{compared to} geophones.  One can install optical fiber for the whole length of the borehole and use it for continuous monitoring of either repetitive active-source seismic measurements (i.e. time-lapse seismic) or for passive monitoring, like for hydraulic fracturing \citep{Bakku2015, Karrenbach2019, Becker2017}. A major benefit of using DAS over goeophones in such settings is that it is not disruptive during the production process.
As for active seismic, early adoptions of the technology were mostly in borehole seismic settings, so for Vertical Seismic Profiling \citep{Mestayer2011,Barberan2012,Daley2013,Mateeva2014,Frignet}. The implementations of DAS in surface seismic have been limited mostly to the estimation of near-surface shear-wave velocities from surface-wave signals with either active or passive sources. Examples of the use of  DAS with \sout{active} passive sources utilizing interferometry techniques for shear-wave velocity inversion are shown in \cite{Dou2017,Ajo-Franklin2017, Tribaldos}. As for active-source implementations, examples include \cite{Song2018, Cole2018}.

For reflection seismic in surface deployment, field tests have been very limited due to the low \sout{decreased} sensitivity of the fiber to broadside waves \citep{Kuvshinov2016}. An indirect approach for \sout{to} obtaining surface recordings of seismic data with DAS was implemented by \cite{Bakulin2017}; their approach involved obtaining reflection data by using multiple shallow upholes. A more direct way to obtain reflection data is to install helically wound fibers near the surface and in a horizontal borehole \citep{Hornman2017, Urosevic2018, Spikes2019}. A more recent approach \sout{application} to enhance the broadside sensitivity is discussed in the work of \cite{White2021}, where multiple fiber configurations and arrangements are used. For multi-component DAS measurements,  theoretical models are proposed for the use of multi-helix configurations \sout{approaches} to retrieve the strain-tensor components \citep{Innanen2017, LimChenNing2018a}. 

In this paper, we discuss a field experiment in which we examine the combined use of straight and helically wound fiber\sout{,} for land seismic measurements. Our focus is on the analysis of multi-mode information obtained from the combined use of straight and helically wound fibers as well as assessing the usefulness of their \sout{the} combined use for reflection seismic.

The paper is structured as follows. In the introduction, we will describe the principles of DAS, how it relates to geophone data, and how it can be modeled. This is followed by describing the field experiment with the set-up and fiber configurations. The obtained results are shown and compared to each other, compared to geophone data and our interpretations are supported by modeling. Some open points are then discussed \sout{Discussion on some open points is then presented} and finally the conclusions are given. 

\subsection{Distributed Acoustic Sensing: Principle and Applications}

A DAS system consists of an interrogation unit or interrogator and an optical fiber, one of its ends being connected to the interrogator.
It is based on the principle of Optical Time Domain Reflectometry (OTDR) where the sensing fiber is injected with a light pulse/signal. As the light propagates and is guided through the fiber, it encounters randomly distributed inhomogeneities infused in the fiber material and it is scattered in all directions. This is what we call the Rayleigh-scattering mechanism, which is an elastic scattering process \sout{where elastic scattering means} meaning that the frequency of the incident light and the scattered light is the same. 
Only back-scattered light is then captured by a photodetector in the interrogator and the phase information of the light is obtained (see Figure \ref{fig:figure1}a). To understand how DAS records a passing seismic wave, a very simplistic illustration is presented in Figure~\ref{fig:figure1}. To properly describe this, two time scales are \sout{should be} considered, namely a fast time $\tau$ and a slow time $t$. The fast time $\tau$ corresponds to the transit time of light within the optical fiber, while \sout{and} the slow time $t$ corresponds to the time of the passing seismic wave. For sake of simplicity, imagine just two scatting points separated by a distance of $x$ in rest position as shown in Figure~\ref{fig:figure1}b. As a wave passes in $t$ time, the relative distance between the scattering points is \sout{then} changed by $\delta x$ and therefore the phase of the back-scatter light is illustrated in Figure \ref{fig:figure1}c which results in the measured strain shown in Figure \ref{fig:figure1}d. That relative change of distance, i.e. $\varepsilon_{xx}= (x+\delta x)/x$ between these scatterers, is proportional to the phase difference $\Delta \Phi$ of the back-scattered signal with the following relationship \citep{Lindsey2020}:
\begin{equation}\label{eq:dphi_exx}
    \Delta \Phi (t, x_i) = \frac{4\pi n L_g \xi}{\lambda} \left(\frac{ x+\delta x}{x}\right), 
\end{equation}
where $n$ is the refractive index of the fiber material, $L_g$ is the length over which the strain is determined, the so-called gauge length, $\xi$ is a scalar factor accommodating the changes in the index of refraction due to stress and $\lambda$ is the wavelength of the source. $\xi$ values vary between 0.79 for pure silica \citep{Schroeder1980} and 0.735 for GeO$_2$-doped fibers \citep{Bertholds1988}. A typical value for $n$ is 1.45 for a wavelength $\lambda$ of 1550 nm \citep{Lindsey2020}. In practice, the measured phase change $\Delta \Phi$ ranges from $-\pi$ to $\pi$. Higher phase values are wrapped and need to be unwrapped first, before retrieving the strain $\varepsilon_{xx} = (x+\delta x)/x$ with expression (\ref{eq:dphi_exx}).

Note that in practice, the majority of DAS systems extract the temporal change of the phase, i.e. $\partial_t  \left(\Delta \Phi\right)$, and therefore the output of the DAS system will be the strain-rate instead of strain where $\partial_t \left(\varepsilon_{xx}\right) \propto \partial_t \left(\Delta \Phi\right)$. Several methods are used to retrieve $\Delta \Phi$, or its time-derivative; these include interferometric approaches using a single pulse \citep{Aa2000,Masoudi2017,Farhadiroushan2010}, a dual-pulse approach \citep{Dakin1990} and heterodyne \sout{hytrodyne} approach \citep{Hartog2012}. More recent advances in DAS include the use of a chirp pulse as a source to allow long-range measurements of as much as 171 km \citep{Waagaard:21}. 

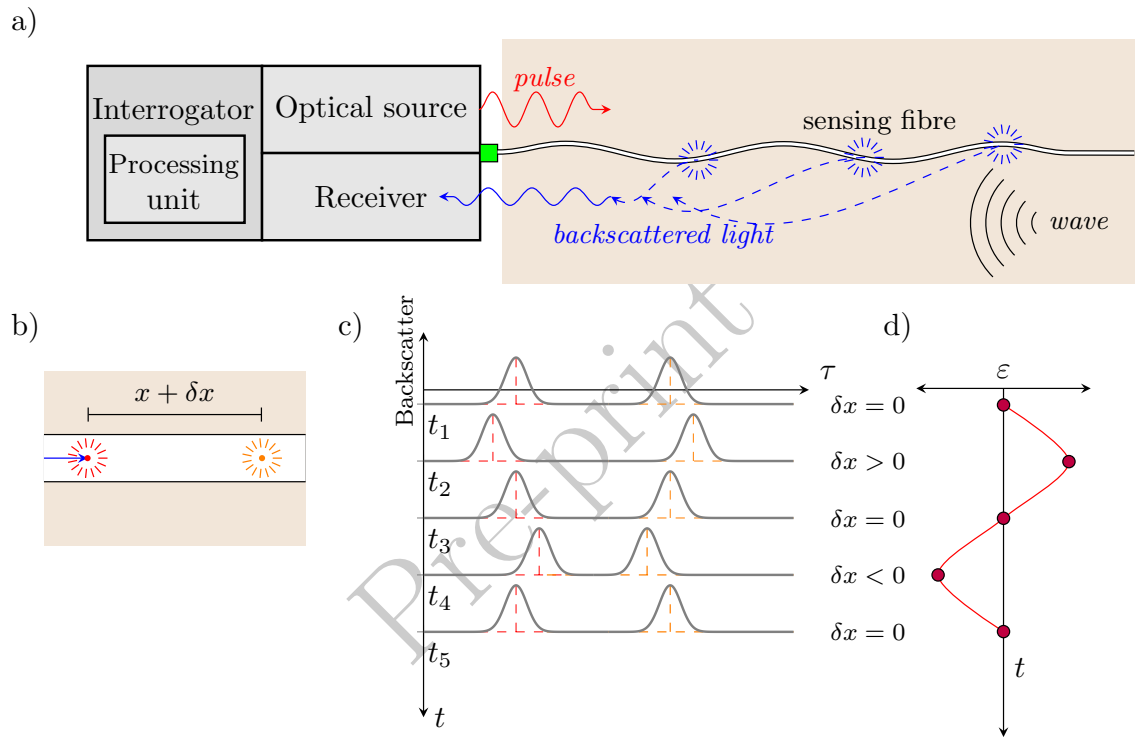
\plot{figure1}{width=\textwidth}{Illustration of DAS principle. a) Main components of a DAS system: an optical source sending a pulse into the fiber, and  back-scattered light to be captured and processed. b) Two scattering points separated by distance $x+\delta x$. c) Perturbation of phase of the back-scattered light from the scattering points in (a), to be translated into a strain measurement as in (d).}

\subsubsection{Gauge length, noise floor and pulse repetition frequency}
Several factors should be taken into consideration to measure the desirable seismic response. Here, we consider the most relevant factors that affect the DAS measurement. The first and most important one is the gauge length $L_g$. $L_g$ is the length over which the strain $\varepsilon_{xx}$ is calculated, as shown in equation \ref{eq:dphi_exx}. So, $\varepsilon_{xx}$ is an average strain along $L_g$. \sout{It is an average strain along a certain length  of the fiber.} Two factors determine a suitable gauge length \citep{Dean2017}, namely sampling the seismic wavefield densely and having a sufficient S/N ratio. For DAS measurement, there is a trade-off between choosing a small gauge length to sample the wavefield densely and decreasing the S/N ratio. The choice is therefore a compromise. Another factor affecting the quality of the DAS measurement is the pulse repetition \sout{pulse-rate} frequency (PRF) of the light source. To minimize the noise/error coming from the phase-unwrapping process, hence decreasing the noise floor, the PRF should be as high as possible but up to a certain maximum $\mbox{PRF}_{max}$ to avoid overlapping with subsequent pulses \citep{Fernandez-Ruiz2019}. The limit $\mbox{PRF}_{max} $ can be expressed as
\begin{equation}\label{eq:PRF}
    \mbox{PRF}_{max} = \frac{c}{2n_g L},
\end{equation}
where $c$ is the speed of light in vacuum, $n_g$ is the group refractive index, and  $L$ is the total length of the fiber in meters. The amount of self-noise in the system coming from the system components is another factor affecting the S/N ratio. 

\subsubsection{Sensitivity of straight and helically wound fibers}
An intrinsic issue with DAS is that it is based on the elongation and contraction in the direction of the optical fiber. When a strain is applied at an angle $\theta$ to that fiber, a strain sensor shows a $\cos^2\theta$-amplitude dependency, as shown by \cite{Benioff1935}. So also in response to a P-wave propagating with angle $\theta$ to the fiber, such a dependency is present. This means that there is a decreased sensitivity to broadside waves. This is especially relevant \sout{the case} in the context of surface seismic if the aim is to measure reflections; this decreased sensitivity may make it difficult to record them on horizontally surface-deployed fiber-optic cables. This is unlike DAS in borehole applications like Vertical Seismic Profiling (VSP) where the reflected upgoing P-wavefield will have a propagation angle almost parallel to the fiber's axis (i.e. $\theta \rightarrow 0$). 

To enhance the broadside sensitivity for surface-deployed cables, fiber shaping to a helix was introduced \citep{Den2012,Kuvshinov2016,Hornman2017}, commonly known as a helically wound fiber (HWF). An illustration of the HWF is shown \ref{fig:figure2}a with two wrapping angles $\alpha$ of $60^\circ$ and $30^\circ$ which are used in this study. Theoretically, \cite{Kuvshinov2016} shows how the wrapping angle $\alpha$ affects the response to P-waves, S-waves and surface waves. One outcome was that shaping the fiber into a helix will decrease its sensitivity to \sout{will act as a filter for} surface waves as well as enhance the fiber's sensitivity to broadside P-wave reflections.

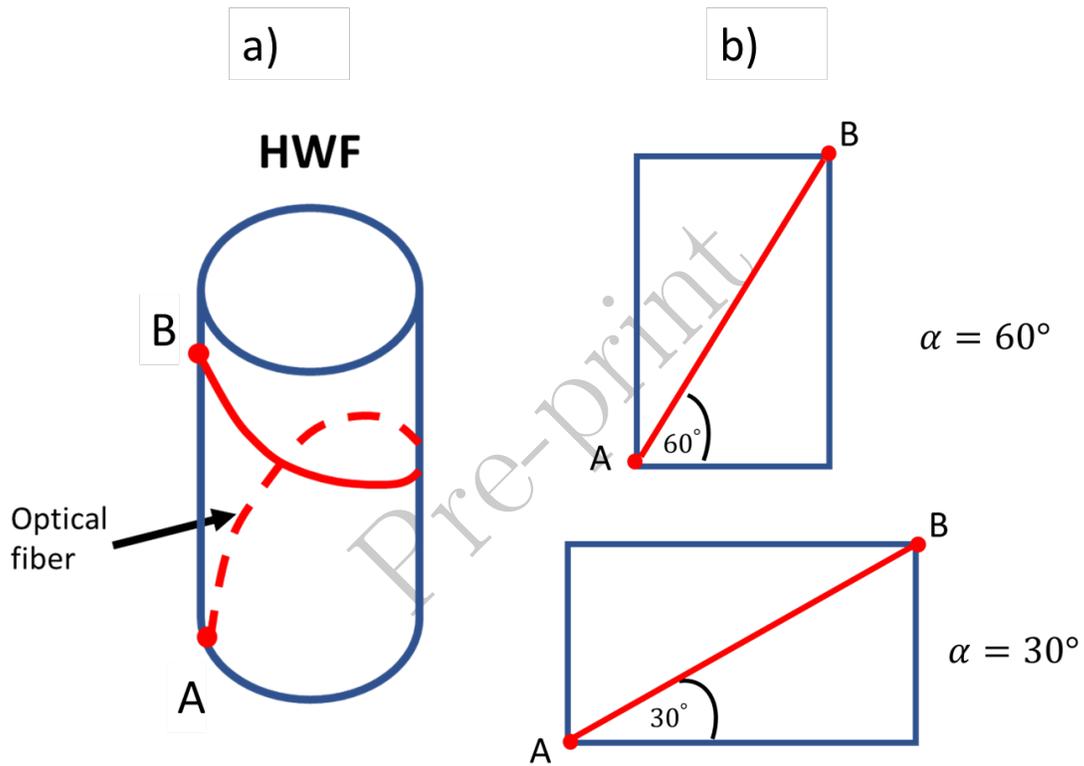
\plot{figure2}{width=\textwidth}{Illustration of the HWF: a) Optical fiber embedded in a cable and b) The wrapping angle $\alpha$. Figure adapted from \cite{Kuvshinov2016}}.

\subsubsection{Modelling helically wound fiber}
To model the helically wound fiber response, we first start by explaining how DAS measurements are related to geophone responses (i.e. particle-velocity measurements). DAS measurements can be equivalently represented as an estimate of the strain rate, i.e. $\dot{\varepsilon}_{xx} = \partial_t \left(\partial_x u_x \right) $, where $u_x$ is the displacement, or as  the spatial derivative in $x$ of the $x$-component of the particle-velocity, i.e. $ \partial_x \left(\partial_t u_x \right) = \partial_x V_x$. We use the latter representation for our modeling. A 2D elastic finite-difference modeling program \citep{Thorbeke2011} will be used to estimate the velocity components and their spatial derivatives.

In the modeling, only \sout{the geometrical} the wrapping angle effect of the fiber is taken into account, and the mechanical properties of the fiber and the embedding cables are not included, thereby assuming the fiber and the embedding cable have the same properties as the surrounding soils/rocks. 
In order to calculate the response of a helically wound fiber, we adapted the approach from \cite{Baird2020} but used the spatial derivative of the velocity as the strain rate. So for a single helix with a wrapping angle of $\alpha$, the measured response will be:
\begin{equation}\label{eq:hwf}
    \dot{\varepsilon}_{H\!W\!F}= \partial_x V_x \sin^2 \alpha + 0.5 \, \partial_y V_y \cos^2 \alpha + 0.5 \, \partial_z V_z \cos^2 \alpha.
\end{equation}
As we are using a 2D elastic finite-difference scheme, expression \ref{eq:hwf} is adapted to be 
\begin{equation}\label{eq:hwf2}
    \dot{\varepsilon}_{H\!W\!F}= \partial_x V_x \sin^2 \alpha + \partial_z V_z \cos^2 \alpha.
\end{equation}
We can see from this equation \sout{the equation above} that for a straight fiber $\alpha = 90^{\circ}$ and the measured strain will become the one of a straight fiber, i.e.  $\dot{\varepsilon}_{S\!F}= \partial_x V_x $. We can combine the measurements of a straight and a helically wound fiber as:
\begin{equation}\label{eq:swfhwf}
 \left( 
     \begin{array}{c}
        \dot{\varepsilon}_{S\!F} \\
        \dot{\varepsilon}_{H\!W\!F}
     \end{array}
  \right)
 =
  \left(
     \begin{array}{cc}
      1 & 0 \\
      \sin^2 \alpha & \cos^2 \alpha
     \end{array}
 \right)
 \left(
    \begin{array}{c}
      \dot{\varepsilon}_{xx} \\
      \dot{\varepsilon}_{zz}
    \end{array}
 \right).
\end{equation}
In this study, we will be using this approach to compute the responses.

\section{Field Set-up}
A field experiment was planned and carried out with the aim to see whether reflections could be recorded on surface-deployed fiber-optic cables, and also to see whether a combination of HWF and SF measurements would provide \sout{obtain} extra information, such as type of motion and type of waves\sout{, type of cable design, etc}.
The field experiment took place on farmland in the province of Groningen, in the north of The Netherlands. A schematic showing the shooting line, geophone line and buried fiber cables is illustrated \sout{shown} in Figure~\ref{fig:figure4}a. The different optical fibers are connected via fusion splicing into two long fibers, as will be described later on.
The interrogators were located in the farmer's shed and connected to the buried cables via a standard single-mode-fiber surface cable of 1 km. Figure~\ref{fig:figure4}b shows the fiber's connection box that is used \sout{. That way, it was easy} to check and resolve issues related to connections of the fibers (e.g. bad splices) and tap testing.  The topsoil is mostly composed of clays, peat and some thin sand layers. %
A shallow borehole next to the buried cable was drilled and the P-wave and S-wave logging tools were used to measure the compressional ($V_p$) and shear ($V_s$) velocities down to around 80 m depth. These logs are shown in Figure~\ref{fig:figure3}. We see in the figure that the near subsurface down to 80 m seems to be varying for the most part around 1600-1800 m/s for $V_p$ and 300-400 m/s for $V_s$ except the layer between 10-20 m where velocities drop to around 1500 m/s and 160 m/s for $V_p$ and $V_s$, respectively.
%
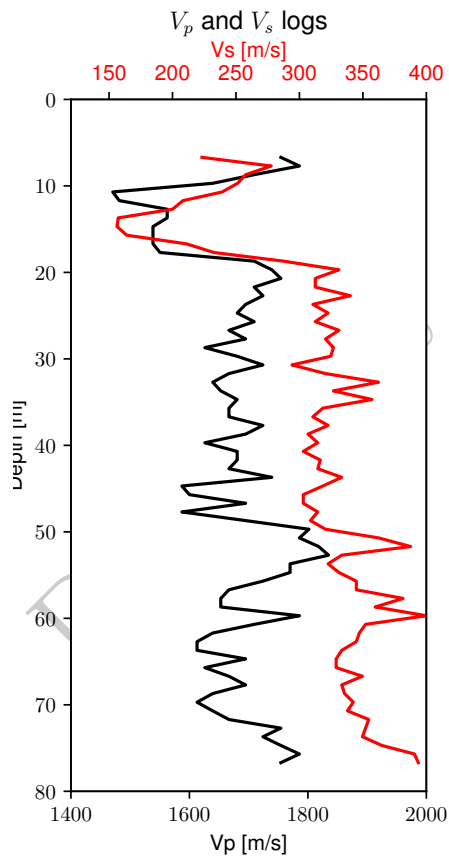
\plot{figure3}{width=0.4\textwidth}{P-wave and S-wave velocity logs of the near-surface.}

The cables are intended for continuous-passive and active time-lapse measurements and were buried 2 m deep, most of the year being below the water table. The cables were laid down by a trenching machine, commonly used for laying out drainage pipes but in this case adapted for our cables, as shown in Figure~\ref{fig:figure4}d. At the time of trenching, we could see that around 2 m depth the soil was quite sandy compared to the clayey/peaty topsoil, which should give us good coupling.  After the cables were laid down in the trench, they were directly covered with the extracted soil by the same machine and afterwards well-compacted by a separate excavator.
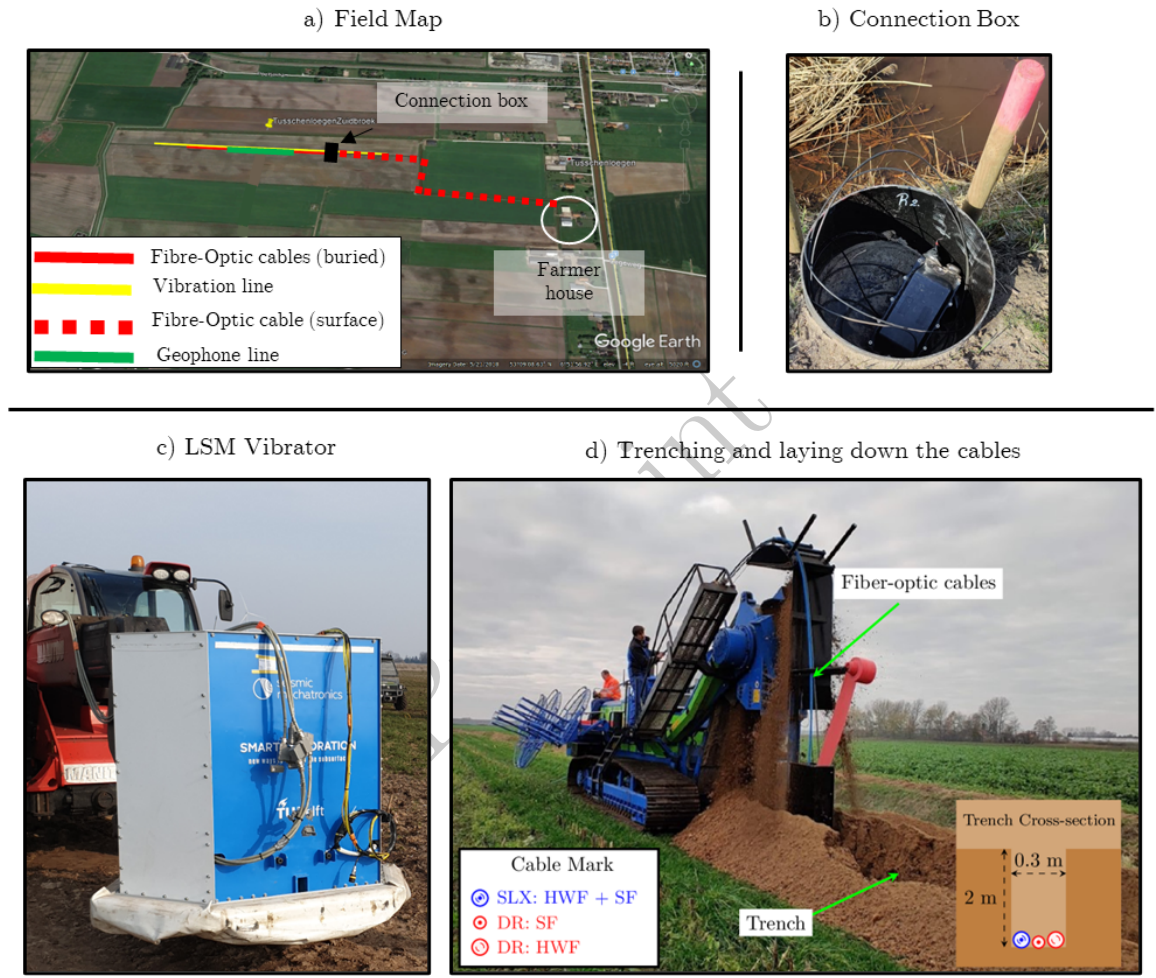
\plot{figure4}{width=\textwidth}{Field experiment set-up and components: a) Field map with  the position of fiber cables (surface and buried), geophone line and vibration line, i.e. shot positions (survey location: 53°9'16.12"N, 6°50'53.99"E), b) Connection box with terminations and splices of cables, \sout{and} c)  Electrically driven seismic vibrator, based on Linear Motor Technology and d) Trenching and burying fiber-optic cables. } 

The source used is an electrically driven seismic vibrator, based on the technology of linear synchronous motors \citep{Noorlandt2015}; it \sout{that} is shown in Figure~\ref{fig:figure4}c. As source sweep, an upsweep of 2-180 Hz with a duration of 12 s was used, with an (extra) listening time of 3 s. As for source position, we shot every 2 m for a 750 m shooting line. \sout{As source positions every 2 m with a length of 750 m was taken} (see Figure~\ref{fig:figure4}). We opted for such a dense spatial source sampling with the aim to satisfy the spatial Nyquist sampling criterion in common-receiver gathers as well as to create a high-fold image. A number of 2 shots per position was used for vertical stacking.
\noindent
\subsubsection{Interrogator systems and fiber-optic cables}
In this field experiment, we have used two DAS interrogators, namely FEBUS A1-R \sout{DAS unit} and iDAS\textsuperscript{TM} v3. We will denote the FEBUS A1-R system as DAS-1 and the iDAS v3 system as DAS-2 for the rest of the article.  DAS-1 is connected to the conventional single-mode (SM) fiber and DAS-2 is connected to \sout{used with} the engineered (i.e. Constellation\textsuperscript{TM}) fiber. It is important to note here that comparing \sout{comparison between} the different DAS units is not the intent of this study, therefore, no direct comparison on the performance of the systems is presented.  On DAS-1, the buried cables contain the following fibers (see Figure  \ref{fig:figure5}): one cable combining helically wound fiber with a wrapping angle of $60^{\circ}$ and a straight fiber, and two separate cables each containing either a straight fiber or a helically wound fiber with a wrapping angle of $30^\circ$. The straight-fiber cable was a steel-armored one, as commonly used in boreholes. For convenience, we number these fibers as we did for the interrogators: SF-2, HWF-2, SF-1 and HWF-1, respectively. The fiber loop order is chosen in this way to minimize optical reflections caused by splices between fibers with different core radii. As for the Constellation\textsuperscript{TM} fibers, we refer to them as SF*-2 and HWF*-2, which are contained in the same cable as SF-2 and HWF-2, unlike SF-1 and HWF-1 that are in separate cables. 
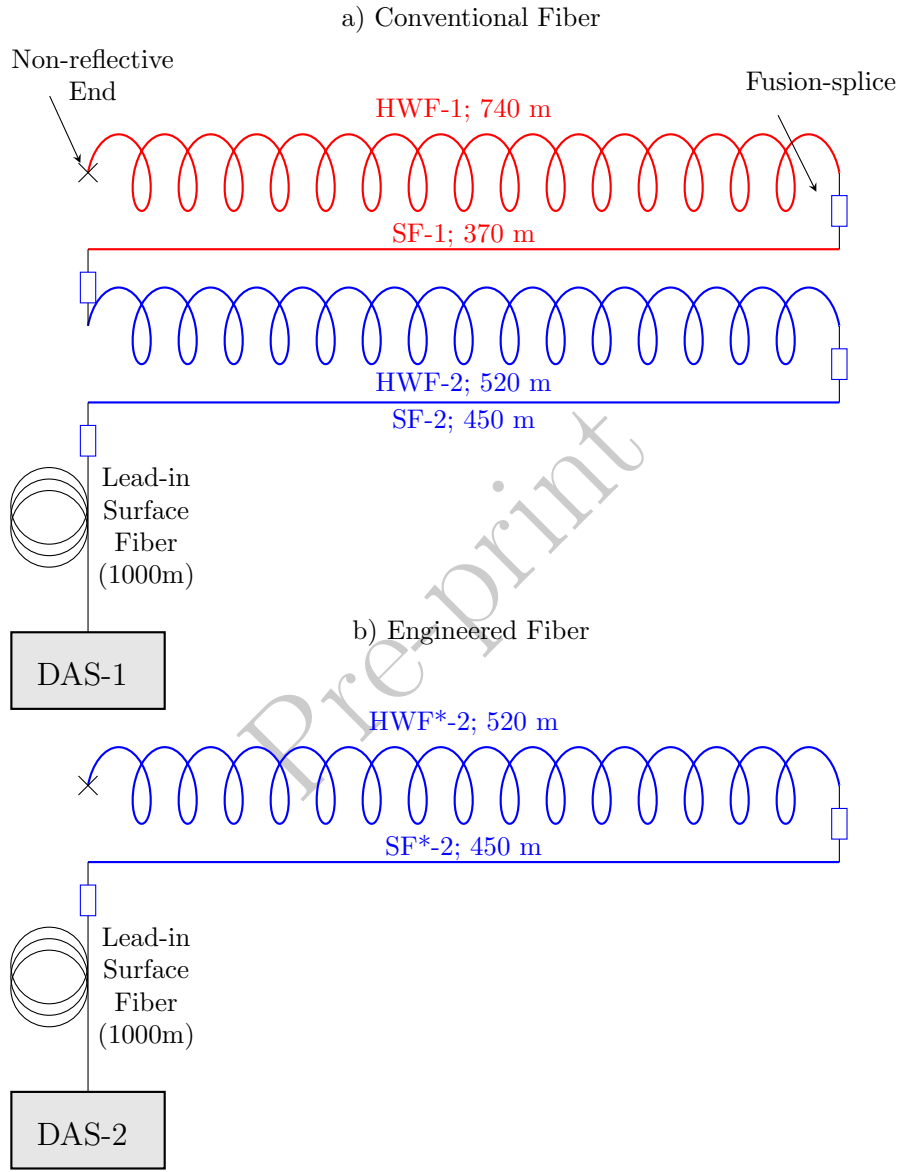
\plot{figure5}{width=12cm}{DAS systems and fiber configurations; a) DAS-1 system connected to conventional fibers and b) DAS-2 system connected to engineered fiber.}
\subsubsection{DAS-recording configuration}
A DAS recording is a more elaborate process to configure than \sout{, compared to} a geophone recording, as it depends on the length of the fiber as well as the amount of optical loss \sout{losses} in the fiber. Each DAS unit was configured separately to have the desired data quality. Information about the acquisition parameters used for DAS is contained in Table \ref{tbl:survey_params}. The most important parameter to note here is the gauge length $L_g$, which was chosen to be 2 m for both systems. Note that even though larger gauge lengths, e.g. 10 m, are expected to give for the most part a better signal-to-noise ratio, our interest was to have proper (Nyquist) sampling of the surface waves to avoid \sout{minimize} spatial aliasing of the total-wavefield recordings. 
\tabl{survey_params}{DAS systems and fiber configurations. (PRF = Pulse Repetition Frequency)}{
\begin{center}
    \begin{tabular}[t]{l>{\raggedright}p{0.25\linewidth}>{\raggedright\arraybackslash}p{0.25\linewidth}}
    \toprule
    
    &\textbf{DAS-1} & \textbf{DAS-2}\\
    \midrule
    \textbf{Gauge length [m]} &2 &2 \\
    \textbf{Output spatial sampling [m]} &1 &1 \\
    \textbf{Fiber type} & Conventional SM  & Engineered SM \\
    \textbf{Total length [m]}&3300 &2000 \\
    \textbf{Trenched length [m]} & 2080 &970 \\
    \textbf{PRF [kHz]} & 10  & 16 \\
    \textbf{Fiber segments} & SF-1, HWF-1, SF-2, HWF-2  &SF*-2, HWF*-2 \\
    \bottomrule
    \end{tabular}
\end{center}
}

\subsubsection{Multi-component geophones}
A portion of the fiber-optic receiver line was \sout{is} also covered with surface-deployed three-component 3-C geophones over a length of 80 m with a spacing of 2 m. \sout{Of the geophones, we} We used the horizontal in-line and the vertical components of the geophones for our comparisons since the former is linked to the strain direction of the straight fibers and the combination of these two components to the signal measured by helically wound fibers. While deploying the geophones, we made sure that they were correctly oriented and properly coupled to the ground. 

\section{Data processing}
A raw measurement included the whole length of the fiber, so different fiber configurations were looped in one long stretch. The raw DAS-1 and DAS-2 data are first down-sampled from 10 kHz and 16 kHz, corresponding to their PRF \sout{PFR} respectively, to a sampling frequency of 500 Hz. Then, the data are correlated with the ground force of the seismic vibrator estimated from several shots. An example of a single correlated shot record is shown in Figure~\ref{fig:figure6}. The beginning of the trench where the fiber gets buried is taken as position 0 m and the fiber before that position corresponds to the surface (extension) cable. The different cables are easily identifiable on the record, as marked by the different colored rectangles below the seismograms, where the white-color rectangles correspond to extra fibers used for connecting the different fibers via fusion splicing. Figure~\ref{fig:figure6} also shows the spectral content of the shot record. Via red arrows, we mark some notable noise sources showing some common-frequency modes. \sout{We can also observe that at a very near offset, as marked by the red box in the figure, the data are contaminated by some high-frequency noise.}
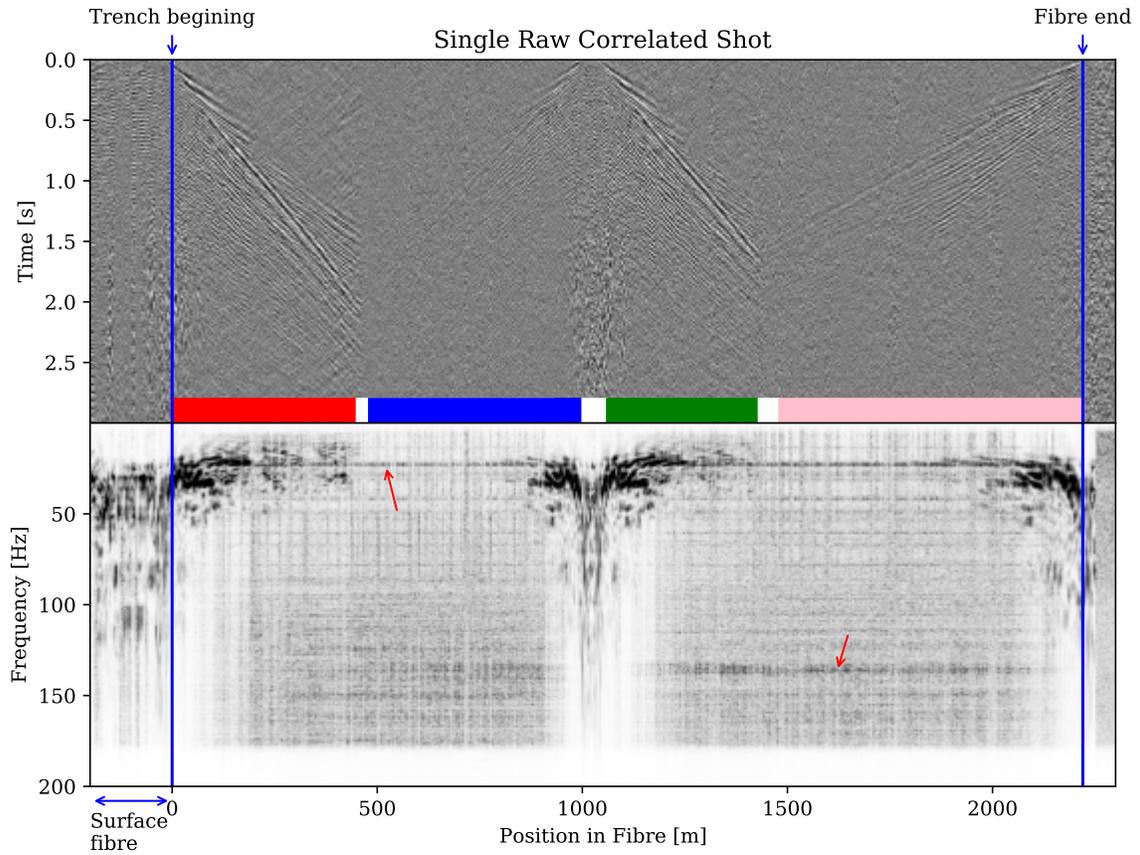
\plot{figure6}{width=\textwidth}{Correlated single raw shot of DAS-1 system in time-distance (top) and frequency-distance domain (bottom). Portions of fiber colored in red, blue, green and pink correspond to fibers SF-2, HWF-2, SF-1 and HWF-1, respectively; white portions are splices.}

Before the survey took place, the fibers were checked with an OTDR device that shows the losses and possible faults along the line. Combining these measurements with the raw data themselves, we were able to separate the different parts of the cables into different data sets with minimal uncertainty in the positions. It is hard to exactly calibrate the distance along the fiber as the spatial resolution is limited by the spatial sampling as well as the gauge length of choice. With our data, the uncertainty in position is estimated at ±1 m.  

Next, processing took place for each data set and the steps are described in the following. We kept the processing quite minimal on purpose as we wanted to minimize processing artifacts. The whole processing flow is shown in Figure~\ref{fig:figure7}. The geometry is set, and vertical stacking is done via a diversity stack. Several noise-removal methods were applied to the data including a trapezoid bandpass filter with corner frequencies 30/40 and 100/120 Hz, FK filter and bottom muting to remove the ground roll as well as some random noise. The data are then sorted into CMP gathers and corrected for NMO with a constant velocity of 1700 m/s, which was determined via the vertical-component geophone data that showed clear P-wave reflections.  Finally, a CMP stack is constructed using RMS-normalzation of each trace before stacking.
 
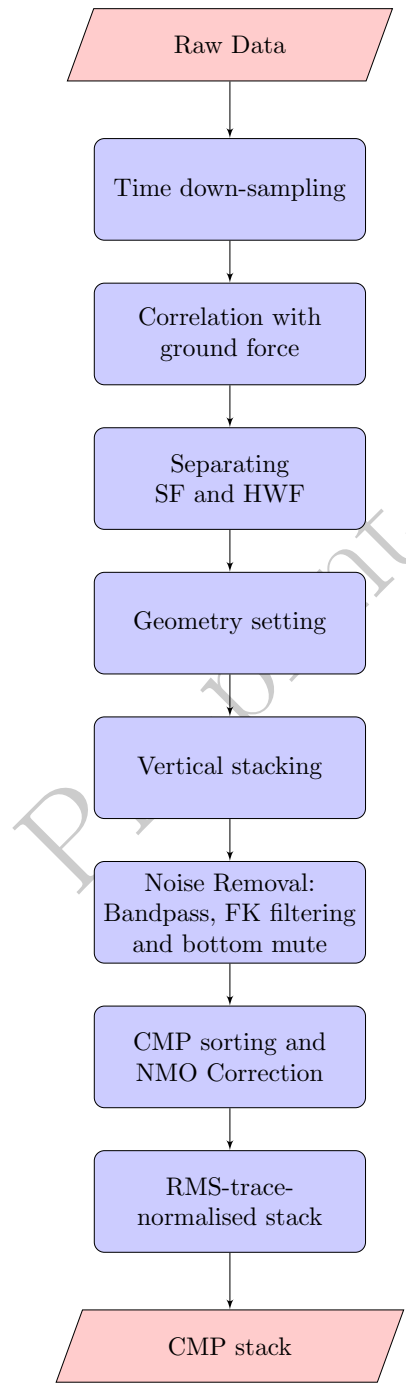
\plot{figure7}{height=0.8\textheight}{Processing flow to produce a CMP stack.}

\section{Results}
In this section, we analyze the \sout{data} pre-stack data, via common-receiver gathers, and the CMP stacks. In the pre-stack domain amplitude and polarity effects in the straight and helically wound fibers are studied, together with reflections observable in these pre-stack data. In the post-stack domain, the results of imaged reflections are discussed.
 
\subsection{Analysis of Common Receiver Gathers}
\subsubsection{Horizontal component of geophone vs. straight fiber DAS}
Here we discuss the comparison in the surface wave signal between the horizontal-component geophone ($H_1$) and the straight-fiber \sout{(SF)} data and  using a common-receiver gather at 400 m and compared to synthetic shots. The synthetics are modeled using a vertical-force source shot at position 400m a 1.5D medium based on the velocity model shown in Figure \ref{fig:figure8}e, which is inspired by the borehole logs (see Figure \ref{fig:figure3}). The density is assumed to have a constant value of 2000 $kg/m^3$.
The source wavelet is a Ricker wavelet with a dominant frequency of 8 Hz. All gathers are filtered with a trapezoid bandpass filter with corner frequencies of 2/4 and 8/10 Hz.

As shown in Figure \ref{fig:figure8}a and \ref{fig:figure8}b, we can see that the SF data is not sensitive to the direction of motion, unlike the horizontal geophones, where the polarity is reversed going from the negative to the positive offsets.  This is expected based on the synthetic examples of the modeled $V_x$ and $\partial_x V_x$ (see Figures \ref{fig:figure8}c and d) which agree \sout{agrees} with recorded geophone and SF data, respectively.

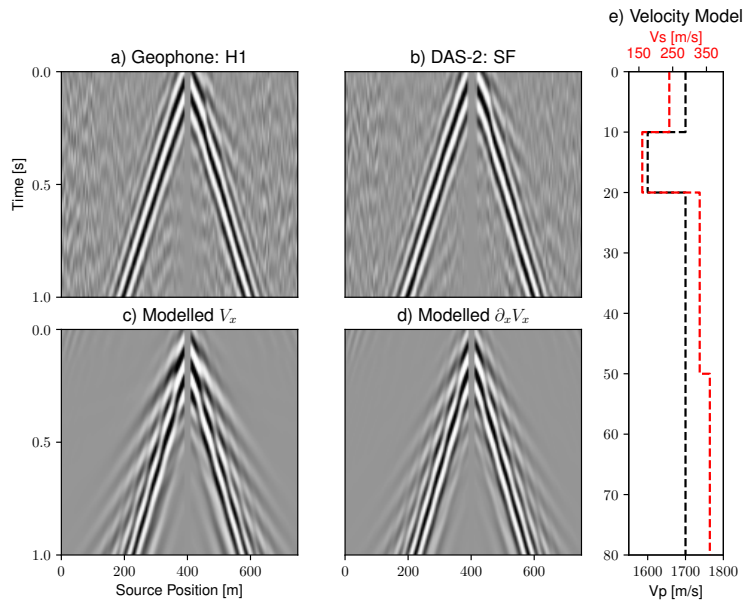
\plot{figure8}{width=10cm}{Common-receiver gathers at 400 m; recorded geophone $H_1$ (a), recorded SF-2 (b), modeled $V_x$ (c), modeled $\partial_x V_x$(d), and the velocity model used to calculate synthetic gathers (e). }

\subsubsection{Response to different fiber geometry}
First, we consider the amplitude differences between the different cables for one receiver position, in this case, 400 m, for all shots. This is shown in Figure \ref{fig:figure9}. When considering the RMS values at \sout{of} that receiver position: even with SF-1 being in a separate (steel-armored) cable and SF-2 being part of a cable that includes both straight and helically wound fibers, the difference in amplitude is minute. This indicates that the cable configuration and its material do not play a significant role in the case of straight fiber.  When the properties of the cable are similar to the ones of the ground around it \citep{Kuvshinov2016,Baird2020}, this can be expected, but it can here be observed that the cable design also does not affect the results. On the other hand, when we compare the amplitudes between SFs and HWFs, the difference is significant as there is an increase of 10-12 dB in favor of the straight fiber as shown in Figure \ref{fig:figure9}. Again, the different types of cables for SF-1, HWF-1 and SF-2/HWF-2 do not alter this, which is notable since the fibers of HWF-1 ($\alpha=30^\circ$) and HWF-2 ($\alpha=60^\circ$) have a different wrapping angle. 

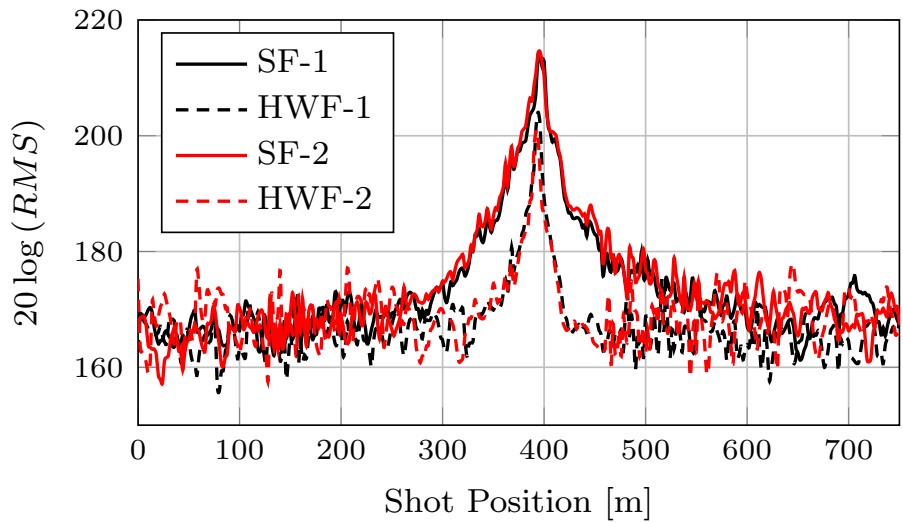
\plot{figure9}{width=12cm}{RMS amplitudes of Common Receiver Gather at 400 m of four different fiber sections connected to the DAS-1 system.}

When looking at the spectra for the different fiber configurations, as shown in Figure~\ref{fig:figure10}, it can be observed that the main differences between SF and HWF occur in the frequency band of some 2-55 Hz, being highlighted in that figure. For data from the area under consideration \sout{such an area}, this is typically the frequency band of the surface waves and S-waves. As \sout{Baird} \cite{Baird2020} noted, an HWF configuration is destructive for S-waves, and therefore also for that component of the surface waves, and here this is confirmed by our observations. The helical shape acts as a (damping) filter for the surface waves and \sout{possible} S-waves. In the band above some 55 Hz, where the information is mainly of P-wave nature (e.g., reflections and head waves), the amplitudes are comparable between the SF and HWF, suggesting that both geometries preserve P-wave information, with a note that these amplitudes are {\it not} enhanced by the HWF configuration, something that would be expected for P-wave reflections.
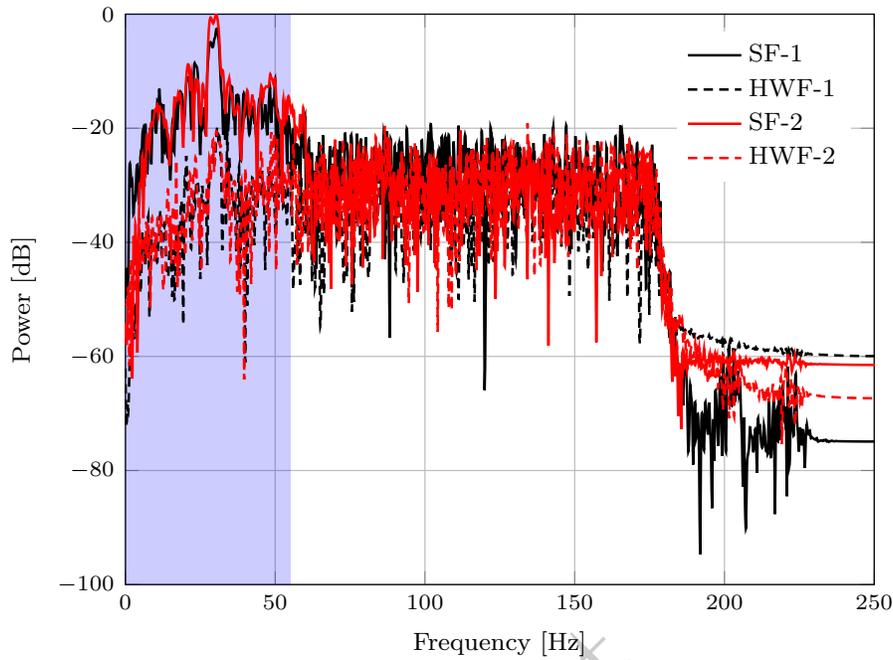
\plot{figure10}{width=12cm}{Power spectra of portion inside the surface-wave cone of CRG at 400 m for different fiber configurations. Band 2-55 Hz is highlighted to indicate the main differences between HWF and SF amplitudes.}

Another type of observation that can be made on the common-receiver gather is the change in phase that can be observed in the surface-wave cone. Figure \ref{fig:figure11} shows the same common-receiver gather (CRG) for the different fibers. We can see that the straight fiber SF-1 and SF-2 give comparable results, as expected. However, change in the polarity of the main surface-wave event around 0.4 s can be observed in the HWF-1 and HWF-2 configurations. These can be attributed to the difference in the wrapping angle. We can see that for HWF-1 ($\alpha=30^\circ$), the polarity is flipped when compared to the straight-fiber signal but with lower amplitude, whereas the polarity is the same for the HWF-2 ($\alpha=60^\circ$) as marked by the red dashed line in Figure \ref{fig:figure11}b.

\plot{figure11}{width=\textwidth}{a) 
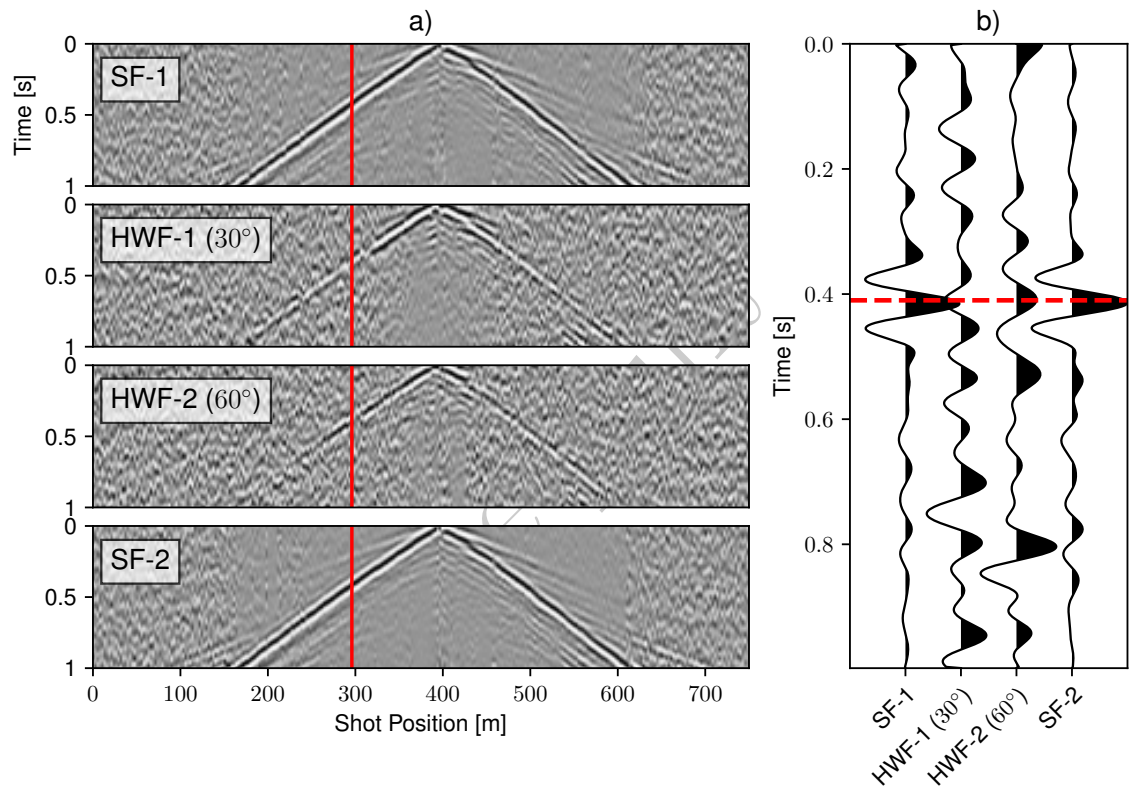
Recorded common receiver gathers at 400 m of fibers: SF-1, HWF-1, HWF-2 and SF-2. b) Traces at shot position 298 m (marked by the red line in a) showing a comparison of signals from the same offset for the different fiber configurations. }

To explain this difference in both amplitude and polarity between the SFs and HWFs, we model their response based on equation \ref{eq:swfhwf} with a simple model shown in Figure~\ref{fig:figure12}d. The  responses are shown in Figure~\ref{fig:figure12}a. Similarly to what was shown in the measured data, we can see an agreement in the changes in polarity and amplitude. The polarity of straight fiber (i.e. Figure \ref{fig:figure12}b) is the same as HWF-2 ($\alpha=60^\circ$), and it is flipped in HWF-1 ($\alpha=30^\circ$) as observed in the measured data. This can be explained by equation \ref{eq:hwf2} where a larger value of $\alpha=60^\circ$ will increase the contribution of $\partial_x V_x$, and will decrease the contribution of $\partial_z V_z$. Also, as shown in Figure \ref{fig:figure12}c), we can see that there is a difference of some 8-10 dB between the straight and helically wound fibers which is close to the measured difference of 10-12 dB as shown in Figure \ref{fig:figure9}. Therefore we attribute the difference in amplitude to the higher sensitivity of the straight fiber to the horizontal component and the lower sensitivity of the vertical component of the surface waves and \sout{possible} S-waves. 
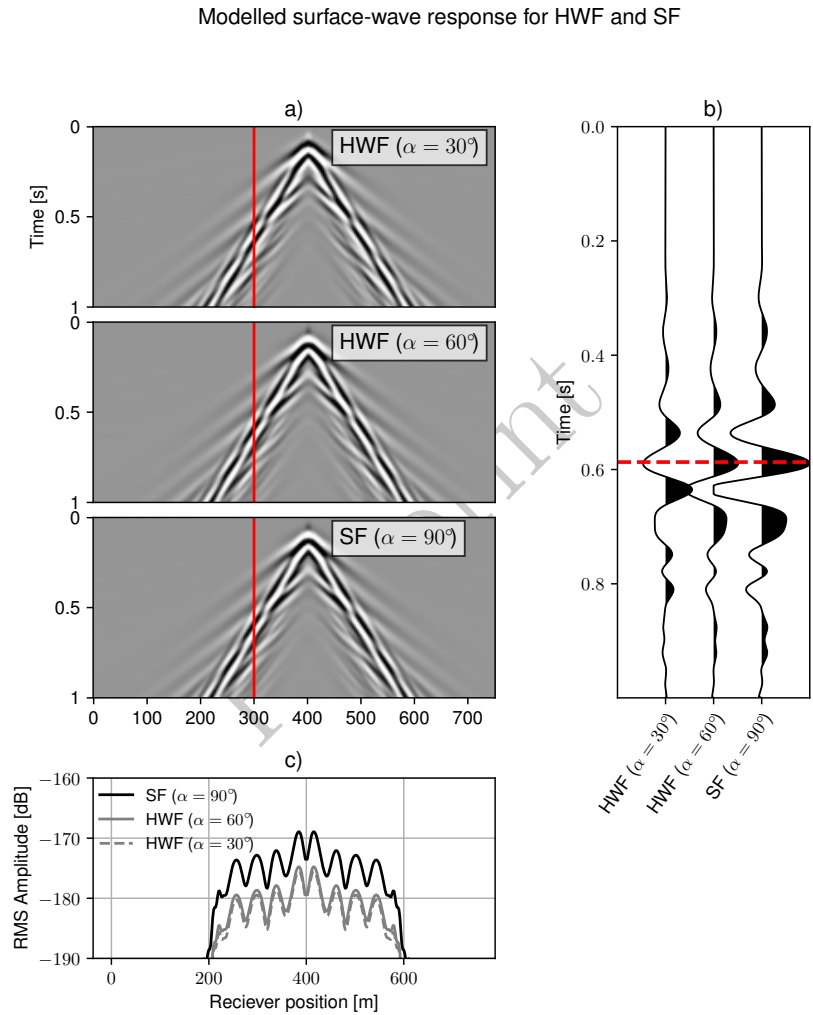
\plot{figure12}{height=0.6\textheight}{a) Modeled surface-wave responses of straight and helically wound fibers with angles and their respective RMS amplitude (c). b) Traces at receiver 300 m showing a comparison between the signals of fibers with different wrapping angles.}

\subsection{Reflections in engineered-fiber recordings}

\subsubsection{Reflections in pre-stack data}
Here we analyze the data collected by the DAS-2 unit connected to the engineered fiber since those data were the best \sout{ones} for this purpose. Figures \ref{fig:figure13}a and \ref{fig:figure13}b show three common-shot gathers of each fiber type at source positions (Sx) 0 m, 10 m and 20 m. For display purposes, the gathers are RMS-normalized and displayed using the same colormap range. The first thing we can observe is that the amplitude is higher in SF*-2 than in \sout{compared to} HWF*-2. In Figure \ref{fig:figure13}c, we look at the RMS values for the time before 0.1 s (marked by the red dashed line in Figures \ref{fig:figure13}a and \ref{fig:figure13}b).  We see that SF*-2 has a higher amplitude overall, as well as higher amplitude variation around the mean RMS value, compared to HWF*-2. Despite SF*-2 having higher amplitude in the early noisy arrivals, we can see the reflections (between the yellow dashed line) more clearly in the SF*-2 data. 

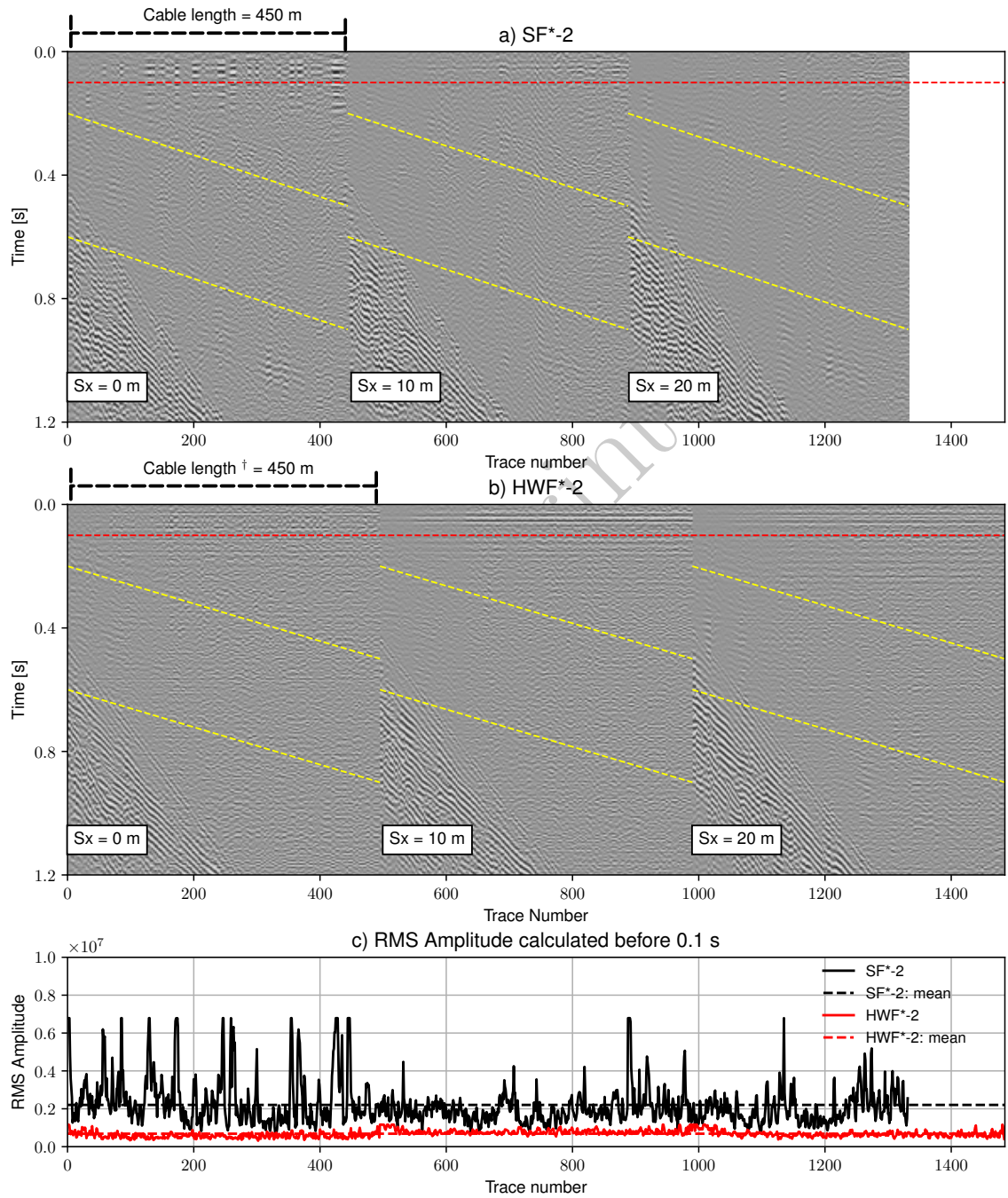
\plot{figure13}{width=1\textwidth}{Common-shot gathers at positions 0, 10 and 20 m for SF*-2 (a) and HWF*-2(b), and c) RMS Amplitude calculated before 0.1 s. } 

 Now we look at the reflections in a common-receiver gather where the reflections seem to be more coherent in this domain. Figures \ref{fig:figure14}a and \ref{fig:figure14}b show a CRG at 173 m for the straight and helically wound fiber, respectively.  As for reflections, we can see three major reflection packages highlighted by the red, yellow and green boxes. When considering the shallow reflections as highlighted by the red box: they are better discernible in the SF*-2 than in the HWF*-2 data. This is somewhat strange since it was expected that a straight fiber should be less \sout{a little} sensitive to a broadside reflection than\sout{,while} the \sout{a} helically wound fiber \sout{would be more sensitive to it,} following \citep{Kuvshinov2016}. The model described therein does not seem to match with \sout{describe} what we observe in the field data.
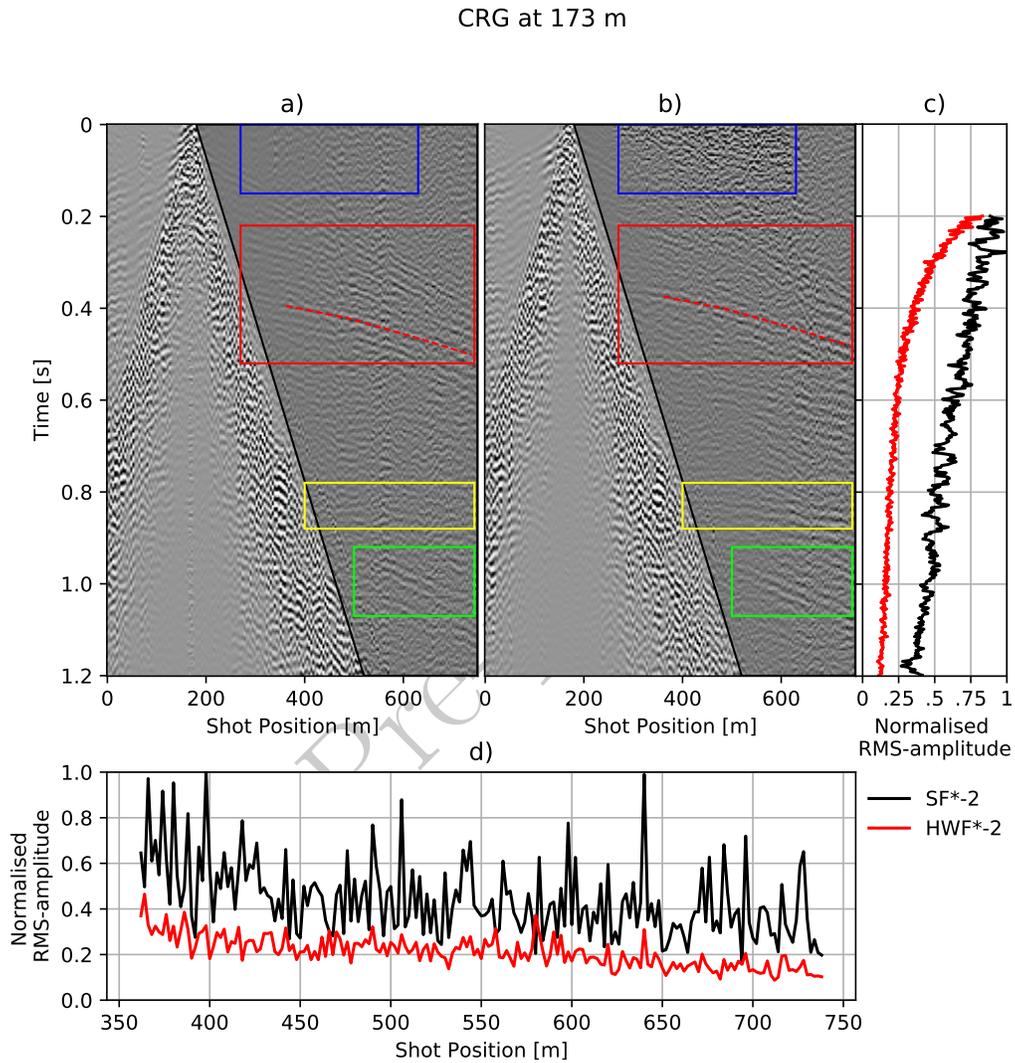
\plot{figure14}{width=0.9\textwidth}{CRG at 173 m of a) SF*-2 and b) HWF*-2 and c) shows normalized RMS amplitude as a function of time, calculated outside the surface-wave cone, as highlighted by the dark grey polygon. d) RMS values calculated along the reflection traced by the dashed red line within a 20 ms window.  Normalized RMS-amplitude for SF*-2 and HWF*-2  calculated as: RMS(SF*-2)/max(RMS($|$SF*-2$|$)) and RMS(HWF*-2)/max(RMS($|$SF*-2$|$)), respectively.}
When following one reflection tracked by the red dashed line in Figures~\ref{fig:figure14}a and \ref{fig:figure14}b, we can see via Figure \ref{fig:figure14}c that the absolute amplitude in SF*-2 is roughly twice as high \sout{much} as the one in HWF*-2. However, we also can see that deeper events as highlighted by the yellow box, are better discernible in HWF*-2 despite the lower amplitude. Lastly, the event highlighted by the green box can be seen equally well in both fiber configurations. Due to the steeper shape of the latter move-out indicating a lower RMS velocity, we interpret this event as a P-S reflection. Again, also these findings do not seem to be completely in agreement with \sout{described by} the model of \cite{Kuvshinov2016}, where the model showed that the normalized strains should be higher in HWF for broadside reflections compared to SF for all propagation angles, but we see that we get higher amplitudes for SF in our measurements. \sout{, where normalized strains are used in the graphs, not absolute ones. }

\subsubsection{Modelling SF and HWF Response to reflection}
To model the reflections, we use a simple model with constant $V_p$ and $V_s$ of 1700 m/s and 360 m/s, respectively and variable density profile to position our reflectors kinematically as desired.  The source is a vertical one at position 173 m. The source wavelet is a Ricker wavelet with a dominant frequency of 30 Hz.  The shots for the straight and helically wound fiber are modeled using eq. \ref{eq:swfhwf} and are shown in Figure \ref{fig:figure15}a and \ref{fig:figure15}b. We can see that the shot for the SF configuration shows the lowest sensitivity to reflections as it \sout{is} mainly contains the horizontal component of the strain-rate tensor.
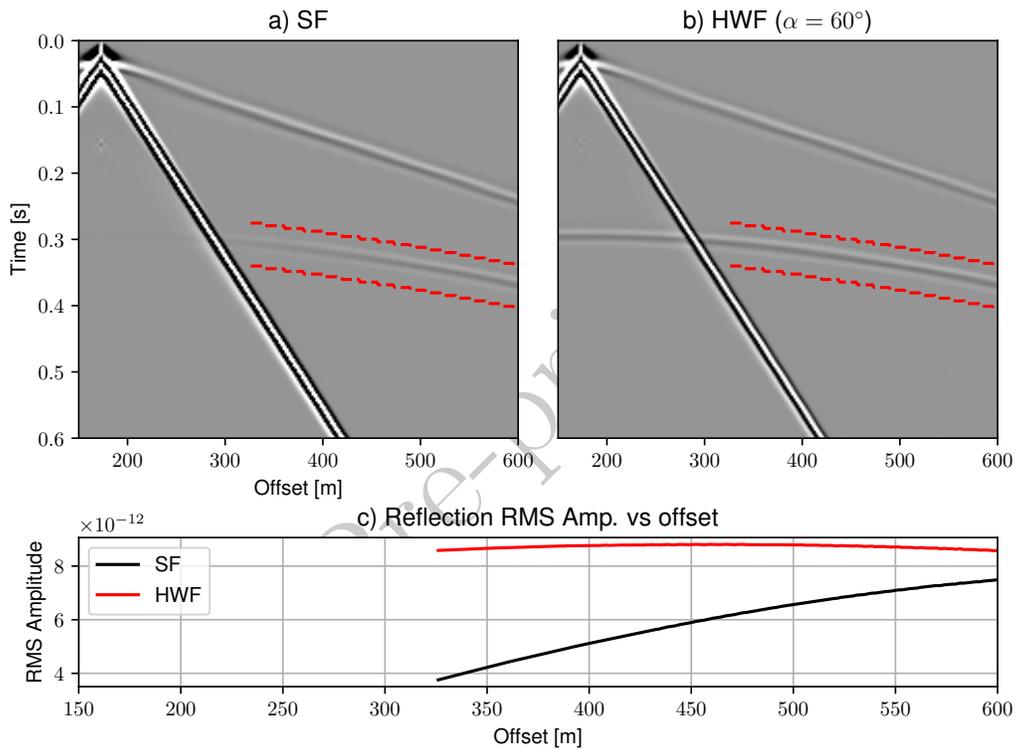
\plot{figure15}{width=0.9\textwidth}{Modeled reflection responses and their RMS amplitudes: shots for SF (a) and HWF with wrapping angle of $60^\circ$ (b) for a density profile with depth range/density values: 0-270/1000 and $>$270/20000 $m$ / $g/m^3$. Both $V_p$ and $V_s$ have constant values of 1700 and 360 $m/s$, respectively.  Dashed red lines highlight analysis windows of 40 ms to calculate RMS-amplitudes as a function of offset for the reflection in (c).}

To analyze this further, we look at the reflections separately as shown in Figure \ref{fig:figure15}a and \ref{fig:figure15}b and calculate the RMS amplitude for every trace at each offset, where windows of 40 ms second around the shallow and deep reflection are highlighted by the dashed lines in the figures. We can see that the HWF ($\alpha=60^\circ$) shows a higher RMS amplitude for the reflection. We can see that for smaller offsets, where the reflected wave is almost vertical, the difference in amplitude is much larger, and for longer offsets, the differences get smaller (see Figure \ref{fig:figure15}c).
\sout{overall. Specifically for the shallow reflection, we see that for smaller offsets, that have steep angles of propagation, the differences in amplitude are much larger and for larger offsets (gentler angles of propagation), the differences get smaller (see Figure \ref{fig:figure15}c). As for deeper reflection RMS amplitudes shown in Figure \ref{fig:figure15}d, we can see that the difference is pretty much constant.} 
This modeling exercise agrees with the model discussed by \sout{what is shown in} \cite{Kuvshinov2016} in terms of HWF having a higher sensitivity to broadside waves, in this case, reflections. However, it is safe to say that this model does not explain the difference in RMS amplitude between SF and HWF observed for the reflections in the measured DAS data (see Figure \ref{fig:figure14}). 
\subsubsection{Reflections in CMP Stack}
We not only analyzed common-receiver gathers for the reflectivity information but also CMP stacks since they would show the quality of images that can be expected. With stacking of NMO-corrected CMP gathers the signal-to-noise ratio is improved substantially, especially because of the relatively high fold due to the small source and receiver spacings. Here we present the CMP stacks derived from the SF*-2 and HWF*-2 data since they gave the best-quality results. 

First, we consider the images produced by SF*-2 and HWF*-2 as shown in Figure \ref{fig:figure16} using the whole fiber length as aperture. Again, it was expected that HWF*-2 would produce the highest-quality result due to the broadside sensitivity, but this was not the case.  For the shallow part, as highlighted by the red box, we can even see that SF*-2 provides better \sout{superior} continuity in the main reflection marked by the red arrow. Still, deeper events, as marked by the yellow arrow, are discernible in the HWF*-2 result, unlike in the SF*-2 one. 

Another way to compare the stacks is to look at the differences using correlation. Traces of CMP numbers 165 to 580 from each stack are windowed, tapered and cross-correlated with each other. Then an average cross-correlation is calculated and shown in Figure~\ref{fig:figure16}c. 
\sout{The average trace of CMP numbers 165 to 580 from each stack is taken, windowed and tapered before the cross-correlation is made. The result is shown in Figure~\ref{fig:figure16}c.} We can see that we have a minimum around $t=0$ indicating that the stacks are out of phase, so they have opposite polarities. 

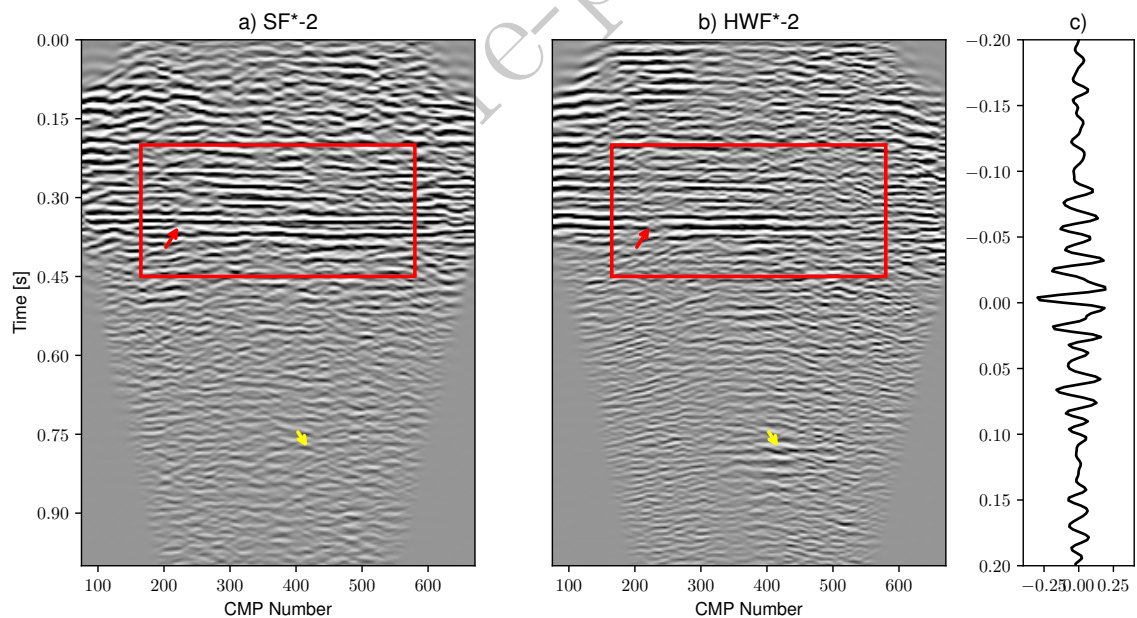
\plot{figure16}{width=\textwidth}{CMP-stack comparison between a) SF*-2 and b) HWF*-2. c) Correlation function between SF*-2 and HWF*-2. The correlation trace in (c) is calculated from the tapered window of 0.2-0.45 seconds over the CMP range of 165-580.}


Second, we examine the continuity and presence of reflections in the SF*-2 data using a comparable fold of coverage as the geophone data. Here only a portion of the fiber is used to create the stack since the same section as the geophone line was taken for comparisons. Next to that, the DAS-receiver line was decimated spatially to the same receiver spacing as for the 3C-geophone line. This comparison is shown in Figures ~\ref{fig:figure17}. 
%
%
Looking at the reference images obtained from the in-line horizontal and vertical \sout{geophones} components, $V_3$ give a significantly better reflection image\sout{, as expected} (see Figures~\ref{fig:figure17}a and b), as expected since most of the reflected energy is nearly vertical. The horizontal geophones gave a worse reflection image since it is mainly sensitive to large(r) propagation angles of reflection. The $V_3$ image is taken as the reference image. 

To compare the geophone data with our DAS measurements, we calculate the horizontal spatial derivatives of the horizontal-component geophone data ($H_1$) and vertical-component geophone data $V_3$  for every position using the expression:
\begin{equation}\label{eq:geop_to_das}
    \partial_x G = \frac{\Delta G}{\Delta x_G} = \frac{G_{i+1} - G_{i}}{\Delta x_G}.
\end{equation}

where $G$ stands for geophone, the index $i$ for the spatial position and it is noted that for our case $\Delta x_G$ is equal to the gauge length $L_g$, i.e., 2 m.

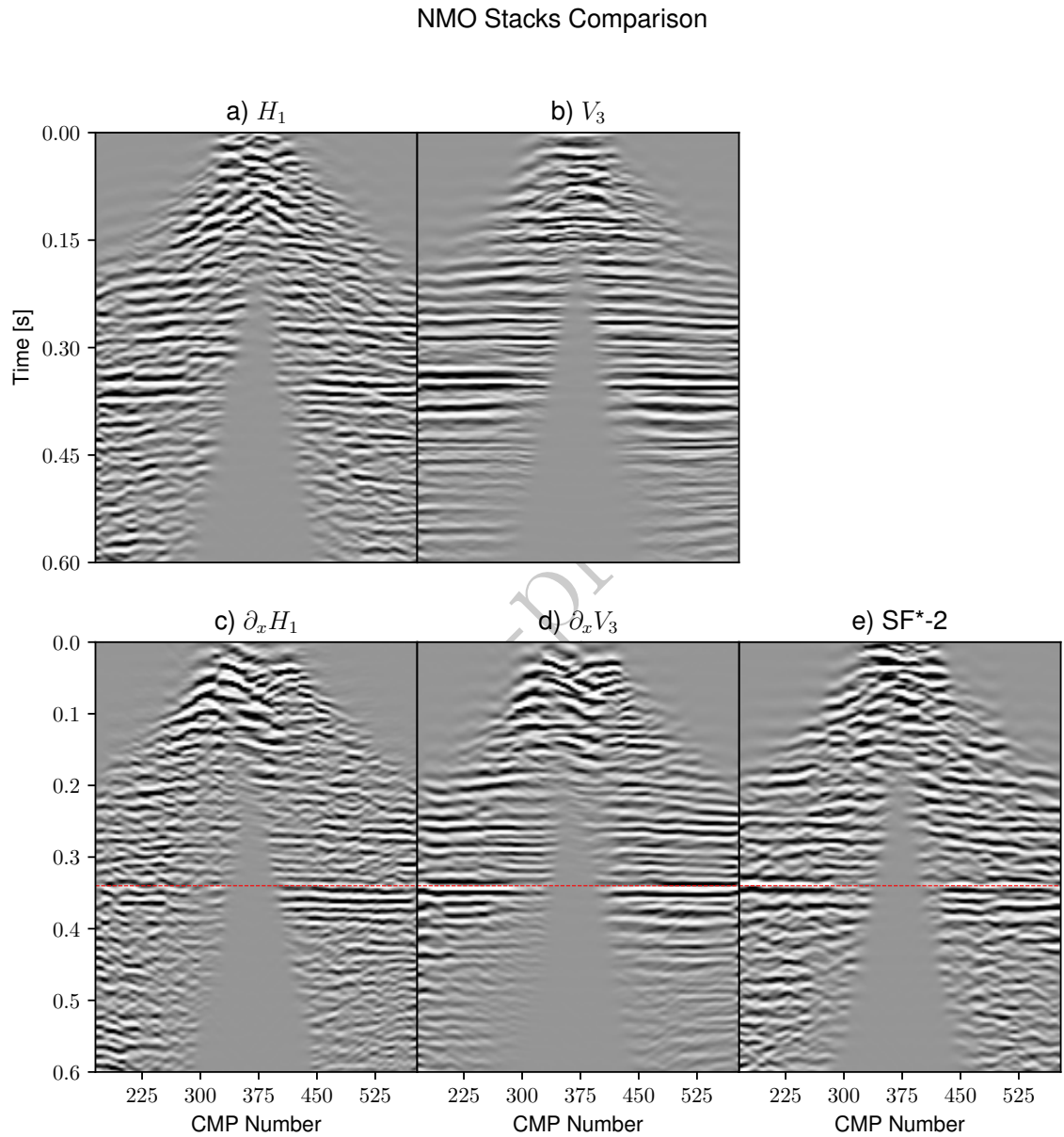
\plot{figure17}{width=\textwidth}{CMP-stack sections for a) Geophone $H_1$, b) Geophone $V_3$, c) (strain-rate) $\partial_x H_1$, d) (strain-rate) $\partial_x V_3$, and e) SF*-2, \sout{multiplied by a factor $-1$ to give same reflection polarity.} All images are RMS-normalized and displayed using the same colormap limits.}

The derived strain-rate responses are shown in Figures~\ref{fig:figure17}c and d next to the SF*-2 shown in Figure~\ref{fig:figure17}e. \sout{It must be noted here that the SF*-2 data were multiplied with a factor $-1$ to obtain the same reflection polarity.} We can see that despite the decreased broadside sensitivity of SF*-2, a reflection image can be obtained even though it is not as good as the one from the vertical geophones. For example, if we look at the reflection at 0.34 s, we can see that the reflection is flatter and more continuous in the $\partial_x V_3$ \sout{compared to} than in the SF*-2. However, the same reflection is better retrieved by the SF*-2 \sout{when compared to} than by the horizontal in-line ($H_1$) geophone data in terms of its continuity. Also, shallow reflections are better shown in SF*-2 data compared to the $H_1$ data, even though both are supposed to be mainly sensitive \sout{mostly} to the horizontal component.
\sout{The strong presence of reflections in SF*-2 compared to $\partial_x H_1$ and the fact that a factor $-1$ was necessary to give the same polarity suggests that there is a Poisson's ratio ($-\nu$) effect in the fiber causing this sensitivity instead of it being merely related to large angles of reflections, so having a large horizontal propagation component near the surface. }

\section{Discussion}
Although we have shown via our field data that reflections are better discernible in SF than in HWF data in the shallow part, the quality should still be improved, especially when compared to the geophone data, and also since we did not see the reflections in all DAS data. The reflectivity in this area is pretty good, also since the water table is very near the surface which helped in detecting reflections. Good-quality systems like the  iDAS-v3 system together with the engineered fiber are definitely needed but systems \sout{that would be having a} with a better signal-to-noise ratios would be even better.


Another issue is that we used helically wound fibers, but it was already suggested by \sout{other author} \cite{Den2012} that sinusoidally shaped \sout{wound} fibers could be used to enhance directivity in a certain broadside direction even more, but this still poses challenges \citep{alhasani2020}.

Our results show that the combined use of helically wound and straight fiber could provide useful insight into the wavefield components, even though a combination has not been done here, e.g. based on inverting equation \ref{eq:swfhwf}; when doing that it did not give satisfactory results. For reflection imaging, we saw that SF*-2 showed better reflection continuity than HWF*-2, while for deeper intervals, reflections are (better) discernible in the HWF*-2 section. This observation could be \sout{better} exploited in a smart combination of these two data sets. 

\section{Conclusions}
In this article, we examined the combined use of straight and helically wound fibers for land surface experiments. We conducted a field experiment in the Groningen area in The Netherlands and showed the behavior of straight and helically wound fibers with different wrapping angles. We analyzed a typical common-receiver gather on its amplitude and some of its phase behavior. We noticed higher amplitudes for surface-wave arrivals in straight-fiber data, in the typical frequency band of those waves \sout{arrivals}, indicating and confirming that the HWF configuration is destructive to surface- and S-wave motion. It was also noticed that the geometry of the cable design, such as cables with a separate (straight or helically wound) fiber or in an integrated fashion (with both straight and helically wound fibers in one cable) had little effect on the observed amplitude behavior. We also confirmed, via our data and a modeling exercise, that the wrapping angle can be such that the surface-wave arrival flips in its polarity. And also that helical winding dampens that signal in quite a predictable fashion.

We \sout{also} saw that for reflection imaging of the engineered fiber data, the straight-fiber data and the helically wound fiber gave similar results \sout{the straight-fiber data gave results similar to helically wound data}. The pre-stack straight-fiber data showed reflection amplitudes of some factor 2 higher than the ones from the helically wound fiber, something that was not expected based on the theoretical models currently in use. In the CMP-stacked data, the straight-fiber section showed more coherent and continuous reflections in the shallow part, despite its decreased broadside sensitivity, while for the deeper reflected events, the helically wound fiber showed slightly better results. It was also found that the stacked SF and HWF data showed opposite polarities for the main reflections. Still, overall, the reflection images are of comparable quality and provide images that are similar to the one obtained from the horizontal derivative of the vertical-component geophone data.

\section{ACKNOWLEDGMENTS}
This research has received funding from the European Research Council (ERC) under the European Union’s Horizon 2020 research and innovation program (grant no. 742703). We would like to thank our technical staff Jens van den Berg, Marc Friebel and Ellen Meijvogel for their assistance. We also extend our thanks to our colleagues Joeri Brackenhoff, Johno IJsseldijk, Marat Ravilov, Menno Buisman, Aydin Shoga, Loes Vogelaar and Travis Hogan for their support in the fieldwork. We also would like to thank Silixa Ltd. and FEBUS Optics for providing support during the fieldwork.

\bibliographystyle{seg}  
\bibliography{main.bib}

\end{document}